 \documentclass[aps,prb,twocolumn,groupedaddress,floatfix]{revtex4}
\usepackage[dvips]{graphicx}

\begin{document}

\title{Screening in semiconductor nanocrystals: \textit{Ab initio} results and Thomas-Fermi theory}
\author{F. Trani}
\email{Fabio.Trani@na.infn.it}
\author{D. Ninno}
\author{G. Cantele}
\author{G. Iadonisi}

\affiliation{Coherentia CNR-INFM and Universit\`a di Napoli ``Federico II'' - Dipartimento di Scienze Fisiche, Complesso Universitario Monte S. Angelo, Via Cintia, I-80126 Napoli, Italy
}

\author{K. Hameeuw}
\affiliation{TFVS, Universiteit Antwerpen, Universiteitsplein 1, B-2610 Antwerpen, Belgium}
\author{E. Degoli}
\author{S. Ossicini}
\affiliation{CNR-INFN-$S^3$ and Dipartimento di Scienze e Metodi dell'Ingegneria, Universit\`a di Modena e Reggio Emilia, via Fogliani, I-42100 Reggio Emilia, Italy}


\begin{abstract}
A first-principles calculation of the impurity screening in Si and Ge nanocrystals is presented. We show that isocoric screening gives results in agreement with both the linear response and the point-charge approximations. Based on the present \textit{ab initio} results, and by comparison with previous calculations, we propose a physical real-space interpretation of the several contributions to the screening.
Combining the Thomas-Fermi theory and simple electrostatics, we show that it is possible to construct a model screening function that has the merit of being of simple physical interpretation. The main point upon which the model is based is that, up to distances of the order of a bond length from the perturbation, the charge response does not depend on the nanocrystal size. We show in a very clear way that the link between the screening at the nanoscale and in the bulk is given by the surface polarization.
A detailed discussion is devoted to the importance of local field effects in the screening. Our first-principles calculations and the Thomas-Fermi theory clearly show that in Si and Ge nanocrystals, local field effects are dominated by surface polarization, which causes a reduction of the screening in going from the bulk down to the nanoscale. Finally, the model screening function is compared with recent state-of-the-art \textit{ab initio} calculations and tested with impurity activation energies.

\end{abstract}


\maketitle

\section{Introduction}

Screening in semiconductor nanocrystals is a fundamental issue, the importance of which is mostly due to the large amount of technological applications inherent in the world of nanostructures.
While screening in bulk semiconductors is a well known and widely investigated subject, \cite{bassani74,pantelides74, resta77, fleszar87,mattausch83} the phenomenon of screening in nanostructures is still not fully understood. The presence of many papers recently published on this subject shows that there is a strong interest in this field.  \cite{cartoixa05,ogut03,franceschetti05,lannoo95,zhou05,delerue03,giustino05}
In particular, it has been shown that the response of a nanostructure to an external field is bulklike well inside the structure. \cite{delerue03, giustino05} This is an interesting feature due to the local nature of bonding in semiconductors. On the other hand, such a behavior is quite different from the trend shown, in both experimental results and theoretical calculations, by the nanocrystal macroscopic dielectric constant,\cite{ wang94, tsu97, lannoo95, trani05} that is shown to increase quite slowly with the size upon going from small nanocrystals to the bulk.
The connection between the microscopic local, bulk-like response and the macroscopic dielectric properties can be understood only through the study of the surface polarization contribution, which strongly influences the nanocrystal properties. 

Preliminary results on this subject have been presented in Ref. \onlinecite{ninno06}. 
In the present paper, we investigate the nature of the screening in semiconductor nanocrystals by performing a detailed microscopic analysis. In the first part of the paper, a first-principles study of shallow impurity screening in nanocrystals is described. The charge density induced by donors and acceptors in Si and Ge nanocrystals is discussed, pointing to the connections to both the point-charge case and the bulk limit.
It is shown that, just like in bulk semiconductors,\cite{pantelides74} the response to isocoric impurities in nanocrystals is similar to that of a point-charge.
A description of the local field effects on the induced charge is given showing that 
the present real-space analysis gives a way to distinguish between surface and bond polarization contributions. In particular, a very interesting comparison with the work of Ref. \onlinecite{fleszar87} allows us to give a physical interpretation of the various contributions to the screening.

In the second part of the paper, a model for the screening is proposed. The well known electrostatic model, based on the image charge method,\cite{lannoo95} is not valid in the neighborhood of the impurity, in that wrong boundary conditions are given for the induced potential. We propose a generalization to semiconductor nanocrystals of the Thomas-Fermi model as originally proposed in Ref. \onlinecite{resta77} for bulk semiconductors.
In order to better appreciate the merits and the shortcomings of the Thomas-Fermi theory, an alternative derivation of the model from the Hohenberg and Kohn theorem \cite{hohenberg64} is described.
In the present model, both the correct limit in 
the neighborhood of the impurity and the surface polarization contributions are taken 
into account from the beginning as boundary conditions of the Poisson equation.
Our results fit well the screening diagonal contribution \cite{fleszar87} for bulk silicon, and excellent agreement with a recent first-principles calculation of the screening function \cite{ogut03} is obtained. Moreover, the model gives a fair prediction of the impurity binding energies.

\section{\textit{Ab initio} results}

We have performed \textit{ab initio} calculations using different kinds of impurities in Si and Ge nanocrystals. 
The study has been based on a plane-wave density-functional theory (DFT) framework.
The calculations have been done with the QUANTUM-ESPRESSO package,\cite{espresso} using the general gradient approximation with ultrasoft pseudopotentials.\cite{vanderbilt90} A vacuum space of at least $6$ \AA{} has been left within the supercell, in order to avoid spurious interactions between a nanocrystal and its replicas. The convergence with respect to the plane-wave basis-set cutoff has been treated with care, and Makov-Payne corrections \cite{makov95} have been added for the computation of the charged nanocrystal total energies.
The nanocrystals have a nearly spherical shape, they are centered on a Si (Ge) atom, and the surface dangling bonds are saturated with hydrogen atoms. The undoped structures have been relaxed, and the optimized geometries used for the doped structures.
In this way we study the screening due to the electronic response to an external perturbation, where fixed ionic positions are considered. The resulting screening does not take into account the contribution due to the ionic displacement. This latter is negligible when covalent semiconductors are considered.

The electron density induced by a donor species $X^{z}$ in the nanocrystal Si$_l$H$_m$ is given by
\begin{equation}
n_{\mathrm{ind}}=n\left[ \left( \mathrm{Si}_{l-1}X\mathrm{H}_{m}\right) ^{z}\right]
-n\left[ \mathrm{Si}_{l}\mathrm{H}_{m}\right],
\label{eq:indchorig}
\end{equation}
where $Z$ is the impurity net charge (atomic units are used throughout this work). A similar expression holds for acceptors and for Ge nanocrystals.
We have focused our study on the spherical average of $n_{\mathrm{ind}}$, and on its integrated density defined as
\begin{equation}
  Q_{\mathrm{ind}}(r) = \int_0^r \bar{n}_{\mathrm{ind}}(x) dx,
\end{equation}
where $\bar{n} _{\mathrm{ind}}(r) = 4\pi r^2 n_{\mathrm{ind}}(r)$.

\begin{figure}
	\includegraphics[width=0.5\textwidth]{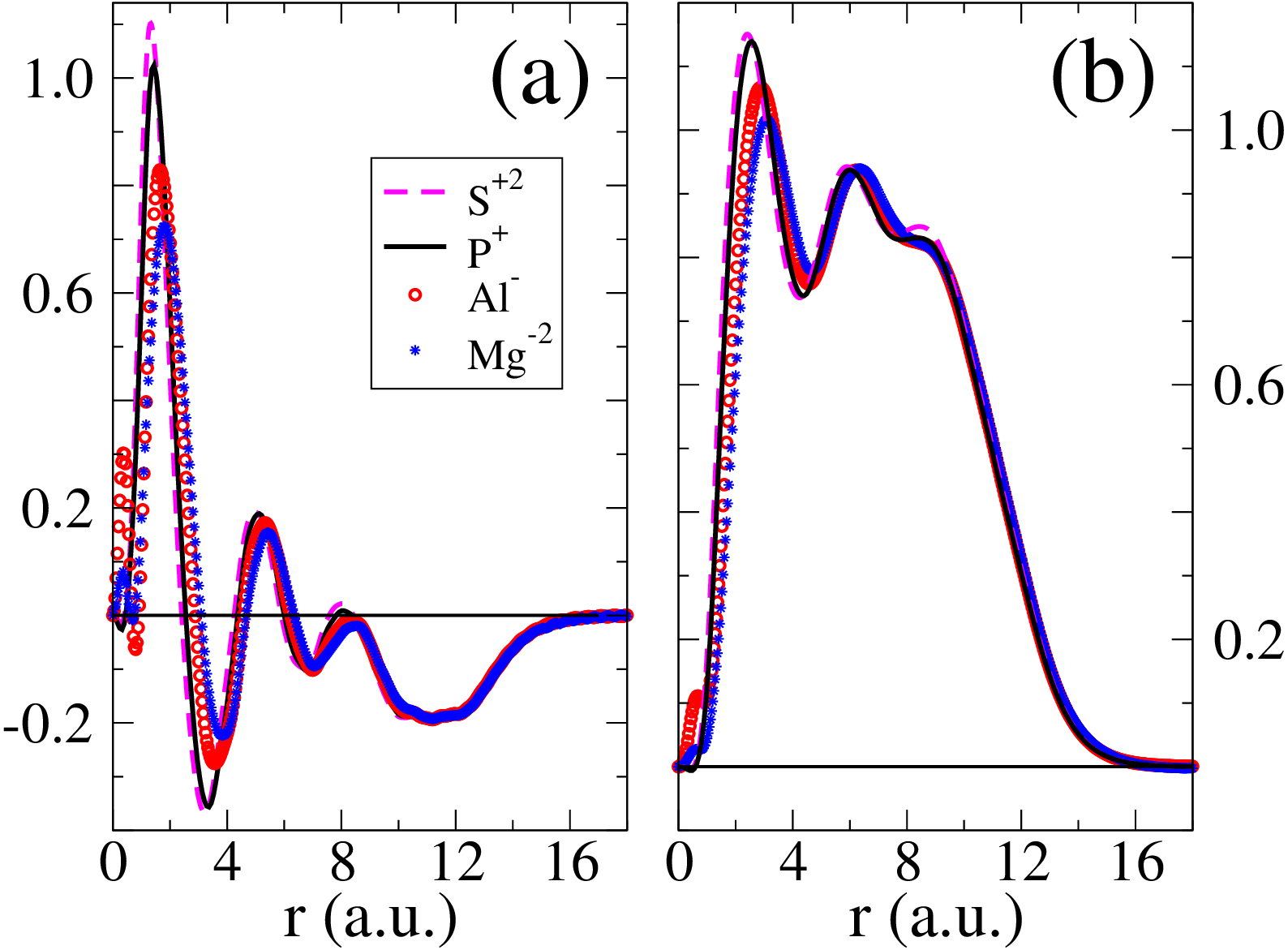}
	\caption{\label{fig:fig1}(Color online) Spherically averaged density $\bar{n}_{\mathrm{ind}}(r)/Z$ [panel (a)] and integrated density $Q_{\mathrm{ind}}(r)/Z$ [panel (b)] induced in Si$_{35}$H$_{36}$ by different doping species: S$^{+2}$ (dashed, pink line) and P$^+$ (solid, black line) donor doping, Al$^-$ (circles, red) and Mg$^{-2}$ (stars, blue) acceptor doping.}	
\end{figure}

In Fig. \ref{fig:fig1}, the results for donor and acceptor isocoric dopants in Si$_{35}$H$_{36}$ are reported. The spherically averaged induced densities $\bar{n}_{\mathrm{ind}}(r)/Z$ [Fig. \ref{fig:fig1}(a)] and the integrated densities $Q_{\mathrm{ind}}(r)/Z$ [Fig. \ref{fig:fig1}(b)] are shown for the donors P$^+$, S$^{+2}$  and acceptors Al$^-$, Mg$^{-2}$. It is worth noticing that the densities induced by P$^+$ and S$^{+2}$ (solid and dashed line, respectively) are almost indistinguishable. This suggests that, at least for donor isocoric impurities, the screening is within a linear-response regime. This was already argued some years ago for the bulk, \cite{pantelides74} and our analysis shows that this is valid for nanocrystals too. From Fig. \ref{fig:fig1} it is seen that, in the case of the acceptors Al$^-$ and Mg$^{-2}$ (circles and stars), apart from a narrow region of space near the impurity (inside the pseudopotential core, the acceptor curves show a nonlinear peak that is absent in the donor case), the overall structure is similar to the donor case. However, while in the region of space after the second main peak from the impurity the curve is very similar to the donor case, the first main peak is lower and pushed away from the impurity site $(r=0)$.
This discrepancy between donor and acceptor doping is in agreement with some results recently published, \cite{franceschetti05} and is related to nonlinearity effects.
The problem of nonlinear screening in bulk semiconductors has been studied with care in the past within a Thomas-Fermi formalism. In Refs. \onlinecite{cornolti78,scarfone84,scarfone85,chandramohan86}, it was shown that there is an asymmetry in the response when positive or negative charged impurities are considered. While for donor screening ($Z=+1,+2$) nonlinear effects are negligible, in the case of acceptor screening the response is quite far from linearity.
As we shall see in the next section, the screening in semiconductors is mostly confined within a screening sphere around the impurity.
The deviation of the radius of such a sphere (screening radius) from the linear case furnishes a realistic check of the linearity of the response. From the Thomas-Fermi theory, an increase of the screening radius of about $0.39$ and $0.68$ a.u. (with respect to the linear case) was observed for $Z=-1$ and $-2$, against a decrease of only $0.16$ and $0.26$ a.u.  found for $Z=1$ and $2$, respectively. \cite{scarfone84,chandramohan86} These results show the same trend as in Fig. \ref{fig:fig1}, confirming our first hypothesis of a good linearity of the response in the donor case.

\begin{figure}	
	\includegraphics[width=0.5\textwidth]{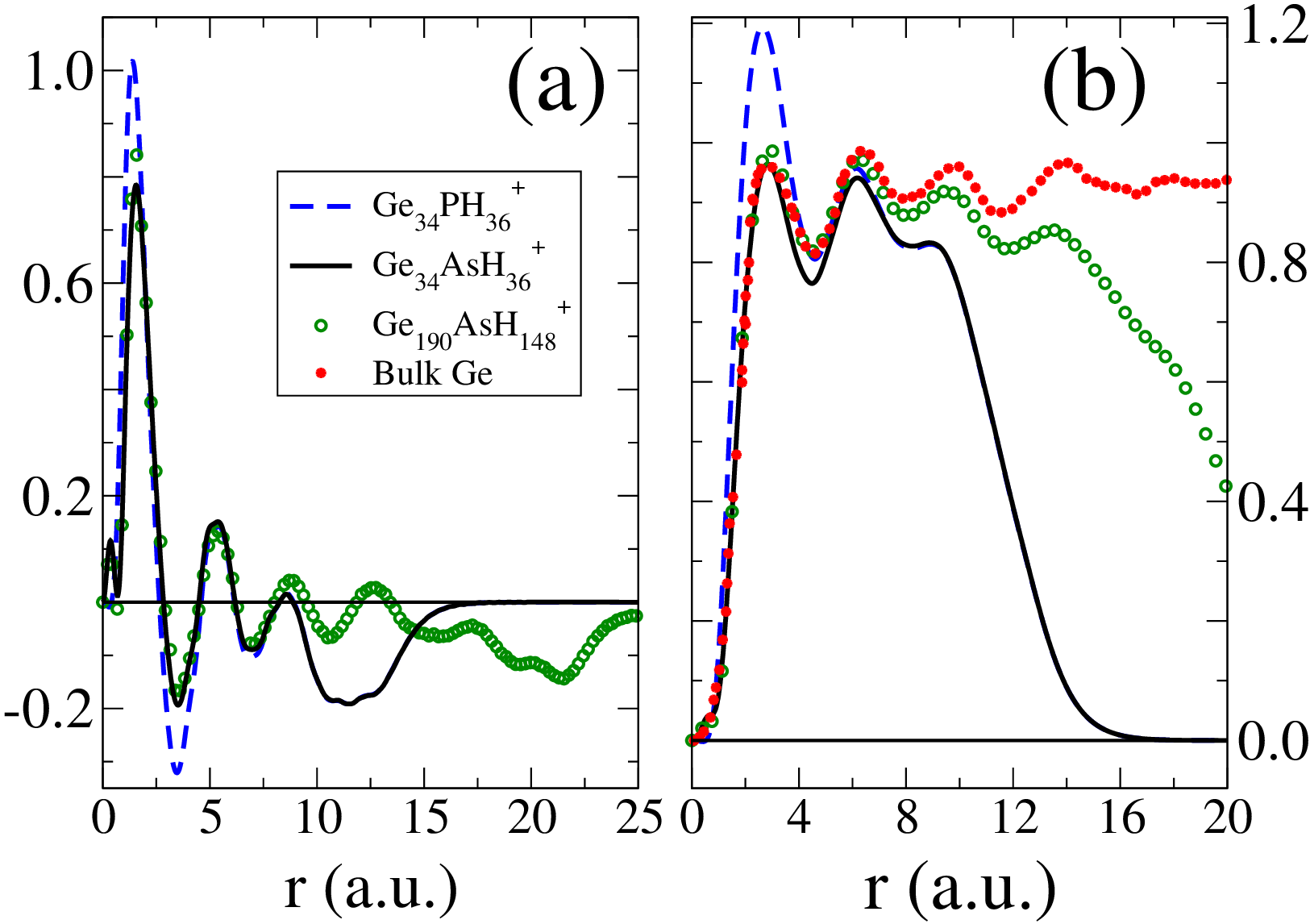}
	\caption{\label{fig:fig2}(Color online) Spherically averaged density $\bar{n}_{\mathrm{ind}}(r)/Z$ [panel (a)] and integrated density $Q_{\mathrm{ind}}(r)/Z$ [panel (b)], induced Ge$_{35}$H$_{36}$ and Ge$_{191}$H$_{148}$ nanocrystals. The results shown are for P$^+$ (dashed, blue line) and As$^+$ (solid, black line) in Ge$_{35}$H$_{36}$, As$^+$ (circles, green) in Ge$_{191}$H$_{148}$. In panel (b) we also report the linear response to a point-charge impurity in bulk Ge from Ref. \onlinecite{fleszar87} (stars, red). }	
\end{figure}

In order to show that the features discussed above for both the induced and integrated density are not peculiar to silicon, we have also studied the case of Ge nanocrystals. In Fig. \ref{fig:fig2}, our results for the induced [panel (a)] and integrated [panel (b)] electron densities are shown for P$^+$ and As$^+$ in Ge$_{35}$H$_{36}$ (dashed and solid line, respectively) and As$^+$ in Ge$_{191}$H$_{148}$ (circles). For comparison, in Fig. \ref{fig:fig2}(b) we also report (stars) a previous \textit{ab initio} result obtained using a truly point-charge impurity for computing the linear-response screening in bulk Ge.\cite{fleszar87}
It is evident that the screening near the impurity is much more pronounced for the nonisocoric P$^+$ impurity than for the isocoric As$^+$. The different core of the dopant atom with respect to the host atom gives rise, in the first case, to the so called central cell corrections, which give a nonlinear contribution. On the other hand, the response to (isocoric) As$^+$ impurity does not induce central cell corrections. It should be pointed out that these corrections have a finite range, being especially pronounced in the region of space before the second main peak in the $Q_{\mathrm{ind}}(r)/Z$ curve. Instead, the electron charge localized on the surface is quite independent of the chemical shift.

Based on Fig. \ref{fig:fig2}, an interesting analysis of the size dependence of the screening can be deduced as well. Indeed, from a comparison between our results for Ge$_{35}$H$_{36}$ and Ge$_{191}$H$_{148}$, we argue that, upon increasing the nanocrystal size, the induced charge close to the impurity rapidly converges to its bulk value. This can be better inferred from panel (b), showing results at small $r$ very close to those calculated in Ref. \onlinecite{fleszar87} for bulk Ge.

From the results illustrated so far, we can reasonably conclude that (i) the isocoric doping is well approximated by a point-charge screening, (ii) the screening is in the range of the linear response regime, (iii) the induced density rapidly converges to the bulk in the region of space close
to the impurity.

Further considerations can be driven from panel (b) of Fig. \ref{fig:fig2}. In Ref. \onlinecite{fleszar87}, it was shown for bulk Ge that the induced charge density [stars in Fig \ref{fig:fig2}(b)] receives contributions from both diagonal and off-diagonal terms of the dielectric matrix in reciprocal space. The diagonal contribution to the integrated density consists in a monotonic increase of $Q_{\mathrm{ind}}(r)/Z$, followed by constant value. Instead, the off-diagonal contribution, related to the so called local field effects, gives a wavy structure to the integrated induced density. This undulating behavior has been discussed by several authors for bulk systems, and arises from the polarization of the bonds.\cite{fleszar87,hybertsen86,mattausch83} Both the diagonal and off-diagonal contributions are retrieved for the nanocrystals, as is clearly seen in Fig. \ref{fig:fig2}, but an additional effect due to the polarization of the surface emerges. It consists of an electron density accumulation around the nanocrystal surface due to the dielectric nature of the structure, as is well known from classical electrostatics. The surface polarization is retrieved from the present \textit{ab initio} formulation and can be seen as a very special kind of local field effect causing the annihilation of $Q_{\mathrm{ind}}(r)/Z$ in correspondence with the nanocrystal boundary [Figs. \ref{fig:fig1}(a) and \ref{fig:fig2}(a)]. We conclude that in a nanocrystal, local field effects (i.e., not diagonal contributions) show up as both the bond and the surface polarization. Nonetheless, while the bond polarization, which is a bulk effect, gives a negligible contribution to the screening and the optical properties, the surface polarization is extremely important. Indeed, it is closely related to the depolarization effects that have been shown to strongly modify the optical properties of low-dimensional structures.\cite{marinopoulos03,bruneval05,bruno05}

\begin{figure}	
	\includegraphics[width=0.5\textwidth]{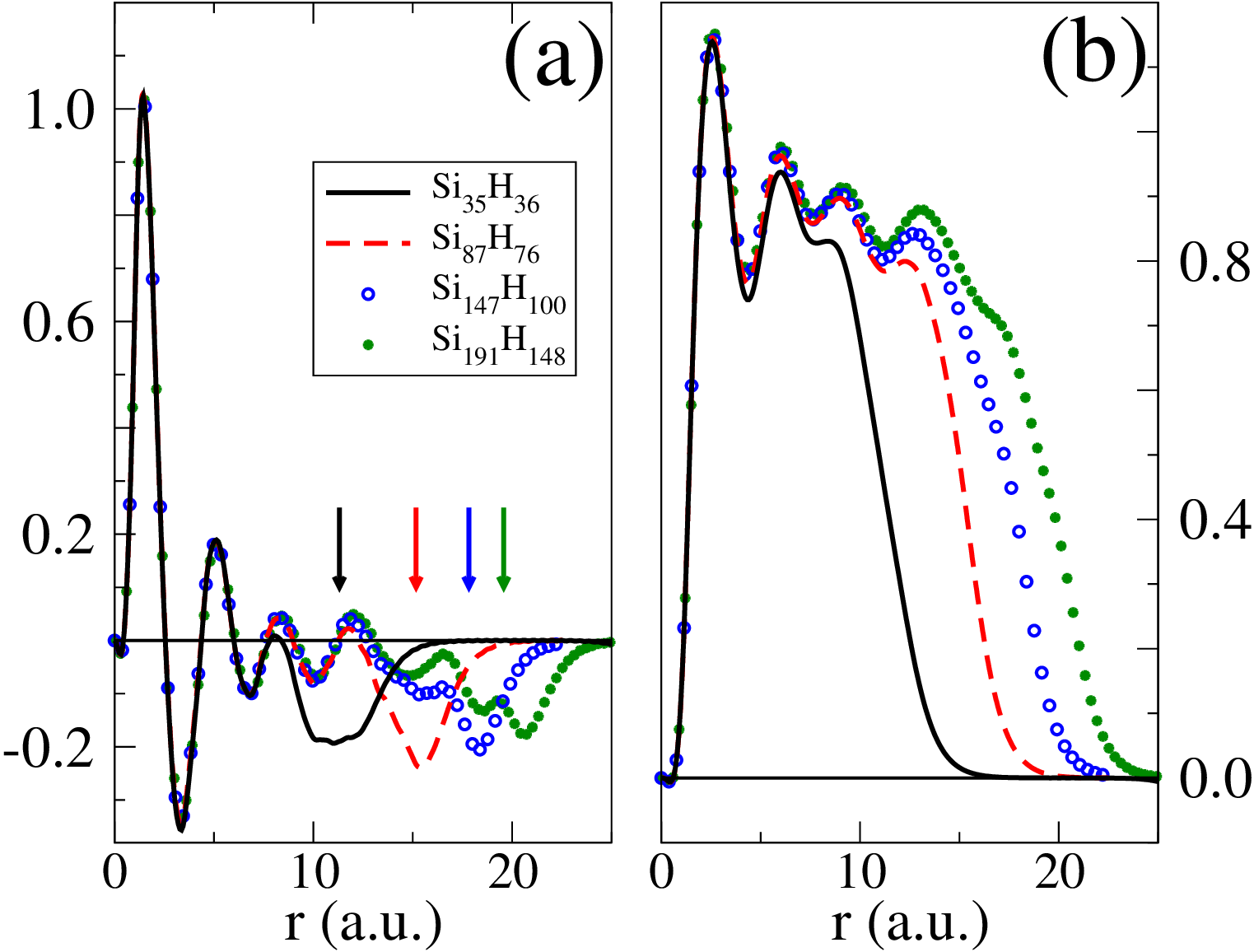}
	\caption{\label{fig:fig3}(Color online) Spherical averaged induced density $\bar{n}_{\mathrm{ind}}(r)$ [panel (a)] and integrated induced density $Q_{\mathrm{ind}}(r)$ [panel (b)], calculated for P$^+$ impurities in Si nanocrystals with increasing size:
	Si$_{35}$H$_{36}$ (solid-black line), Si$_{87}$H$_{76}$ (dashed-red line), Si$_{147}$H$_{100}$ (circles-blue) and Si$_{191}$H$_{148}$ (stars-green). 
	The full arrows point to the nanocrystal radius (see text for definition). }	
\end{figure}

The size dependence of the screening can be inferred from Fig. \ref{fig:fig3}, where the spherical average of the induced density [panel a] and the integrated density [panel b] are shown for a set of P$^+$-doped silicon nanocrystals with increasing radii.
It comes out that up to a distance of a few a.u. from the impurity, both the induced and the integrated densities are almost independent of the nanocrystal size. In order to be more quantitative, it is necessary to define the nanocrystal radius.
We do not follow the standard route, that is, the radius of sphere whose volume equals the product of the volume per atom in the bulk times
the number of, say, silicon atoms. Indeed, the shortcoming of this definition is that the hydrogens passivating the surface are left out. Moreover, in a screening problem the fundamental objects are the electrons, including those participating in the bonds near the surface. On the basis of these arguments, we define the nanocrystal radius through the equation $n_0 4\pi R^3 / 3  = N$, where $n_0$ is the bulk valence electron density and $N=4l+m$ is the total number of valence electrons in Si$_l$H$_m$. The nanocrystal radii calculated in this way are shown in Fig. \ref{fig:fig3}
as arrows. They have slightly higher values than the conventional radii. For instance, in the case of Si$_{35}$H$_{36}$, we retrieve $11.3$ a.u. against the conventional value of $10.4$ a.u.. But, as it is seen from Fig. \ref{fig:fig3}(a), with the present definition all the radii lie almost exactly in the middle of the surface charge density. As a final observation, we stress that despite the undulation in both the induced and integrated densities, the total induced charge around the impurity is almost independent of the nanocrystal radius. Although, at this level, this concept may not appear to be well-defined, we shall see in the next section how it allows the construction of a Thomas-Fermi model for the screening.

\section{The Thomas-Fermi model}
As we have seen in the previous section, the density induced by a point-charge impurity in semiconductor nanocrystals consists of several contributions. The first contribution is due to the reciprocal-space diagonal part of the dielectric matrix, the second contribution comes from the Si-Si bond polarization, and the third is from the nanocrystal surface polarization.
In this section, we illustrate a Thomas-Fermi model, which reproduces with great accuracy both the diagonal and surface polarization contributions.

\subsection{Classical electrostatics}
As a starting step, we briefly describe a simple electrostatic model that can be used as a first approximation in studying point-charge screening in semiconductor nanocrystals.
The basic assumption is that, as a response to a point-charge potential $v_{\mathrm{ext}}(r) = -Z/r$, an electron polarization charge $-Q$ is induced around the impurity. On the other hand, since the total induced charge must integrate to zero, a compensating charge $Q$, assumed as uniformly distributed on the nanocrystal surface, must be introduced. The resulting potential energy is therefore due to a total charge $Z-Q$ localized around the impurity, and to a charge $Q$ distributed on the surface.
From electrostatics, the potential energy for an electron is 
\begin{equation}
v_{c}(r)=\left\{ 
\begin{array}{c}
-\frac{Z-Q}{r}-\frac{Q}{R},\text{ \ }0<r\leq R \\  
-\frac{Z}{r},\text{ \ \ \ \ \ \ \ }r\geq R,%
\end{array}
\right. \text{ \ }  
\label{eq:vc}
\end{equation}
where $R$ is the nanocrystal radius, as defined in the previous section. 

This electrostatic model is supported by the DFT results described in the previous section.
Indeed, Fig.\ref{fig:fig3} (a) shows that the induced electron density has two main contributions concentrated around the impurity and across the surface.
A basic hypothesis for the model is that the total charge $Q$ is independent of the nanocrystal size. This is supported by the fact that, as shown in Fig. \ref{fig:fig3}, the induced density near the impurity rapidly converges to the bulk value.
As is shown below, the use of the bulk value for $Q$ is a good approximation for nanocrystals.

However, this simple model is not accurate in the description of the screened potential close to the impurity site.
Indeed, one can easily see that unphysical potential values are predicted in the limit $r\to 0$. 
In order to correct the model, we follow Ref. \onlinecite{resta77}, in which a Thomas-Fermi description for bulk semiconductors is given. The basic point is the concept of incomplete screening occurring in a semiconductor. This means that the charge is induced only inside a screening sphere of radius $R_s$. Outside the screening sphere, the system behaves as a classical dielectric medium having a static dielectric constant $\varepsilon_s$.
In a semiconductor, the total induced charge $Q$ is a finite fraction of the whole external charge $Z$ introduced with the impurity. It is well known that the ratio $Q/Z$ closely depends on the static dielectric constant through the relation \cite{resta77}
\begin{equation}
  \label{eq:Qvseps0}
  \frac{Q}{Z} = 1 - \frac{1}{\varepsilon _s }.
\end{equation}
The larger the static dielectric constant is, the closer to unit is the ratio $Q/Z$.
The induced potential and the spatial dielectric function are obtained solving the Thomas-Fermi equations with appropriate boundary conditions, where the screening radius is derived self-consistently. \cite{resta77} 

The Thomas-Fermi theory establishes the linear relation between the induced electron density $n_{\mathrm{ind}}$ and the screened potential  $v_{\mathrm{scr}}=v_{\mathrm{ext}} + v_{\mathrm{ind}}$ inside the screening sphere,
\begin{equation}
\label{eq:TFcondition}
   n_{\mathrm{ind}} (r) = -\frac{q^2}{4\pi} \left[ v_{\mathrm{scr}}(r) - \delta \mu \right],
\end{equation}
where $\delta \mu$ (the chemical potential) is a constant to be determined with the boundary conditions and $q$ is a multiplicative constant related to the nanocrystal average valence electron density.\cite{resta77}
Outside the screening sphere, the induced charge is zero and the screened  potential matches the classical expression $v_c$ given in Eq. (\ref{eq:vc}). The continuity of the induced density, which inside the screening sphere is given by Eq. (\ref{eq:TFcondition}), at $r=R_s$  implies  [$n_{\mathrm{ind}}(r) = 0$ if $r \geq R_s$] that
\begin{equation}
\delta \mu=v_c(R_S).
\label{mudef}
\end{equation}

\subsection{Derivation of the Thomas-Fermi theory}
Before going to the application of the Thomas-Fermi theory to a nanocrystal, it is of some interest to rederive Eq. (\ref{eq:TFcondition}) within the DFT framework. From the Hohenberg and Kohn theorem, it is known that the total energy of an electron system, written as a functional of the electron density, can be decomposed into several contributions,
\begin{equation}
 E\left[ n \right] = T _s \left[ n \right] + J \left[ n \right] + E_{\mathrm{xc}} \left[ n \right] + V \left[ n \right],
\end{equation}
where $T_s$ is the noninteracting kinetic energy, $V$ is the total one-particle potential energy
\begin{equation}
   V\left[n\right] = \int d\mathbf{r}\, n(\mathbf{r})v(\mathbf{r}),
\end{equation}
$J$ is the classical Coulomb energy
\begin{equation}
   J\left[n\right] = \frac{1}{2}\int d\mathbf{r}\,d\mathbf{r'}\frac{n(\mathbf{r}) n(\mathbf{r'})}{\left|\mathbf{r-r'}\right|},
\end{equation}
and $E_{\mathrm{xc}}$ is the exchange-correlation energy density functional.\cite{hohenberg64,parr89} 
From the variational principle, with the constraint that the integrated density gives the total number of electrons $N$, we have the stationary condition for the ground-state density $n_{\mathrm{gs}}(\mathbf{r})$,
\begin{equation}
  \left.\frac{\delta E}{\delta n} \right| _{n_{\mathrm{gs}}} = \mu.
\end{equation}
The variable $\mu$ has been shown to be the system chemical potential.\cite{hohenberg64,parr89,liu04} 
In order to study the screening, we write down the Hohenberg and Kohn equations, for both the unperturbed (labeled with 0) and the perturbed (labeled with 1) systems
\begin{eqnarray}
   \label{eq:HKimpert}
   \left. \frac{\delta T _s  [n]}{\delta n (\mathbf{r})} \right| _{n_0}+ v_0 (\mathbf{r}) + v_H [n_0 (\mathbf{r})] +  v_{\mathrm{xc}} [n_0(\mathbf{r})]&=& \mu_0,\\
   \label{eq:HKpert}
   \left. \frac{\delta T _s  [n]}{\delta n (\mathbf{r})} \right| _{n_1}+ v_1 (\mathbf{r}) + v_H [n_1 (\mathbf{r})] + v_{\mathrm{xc}} [n_1(\mathbf{r})] &=& \mu_1.   
\end{eqnarray}
Here, $v$ is the one-electron potential, $v_H$ is the Hartree potential
\begin{equation}
  v _H \left[n(\mathbf{r})\right] = \int d\mathbf{r'} \frac{n(\mathbf{r'})}{\left|\mathbf{r'-r}\right|},
\end{equation}
and $v_{\mathrm{xc}}$ is the exchange-correlation potential.
By subtracting Eq. (\ref{eq:HKimpert}) from Eq. (\ref{eq:HKpert}), and assuming linear response, we obtain an expression relating the induced density $n_{\mathrm{ind}} = n_1 - n_0$ with the screened impurity potential $v_{\mathrm{scr}}$. This potential is defined as the sum of the external perturbation $v _{\mathrm{ext}} = v_1 - v_0$, the induced electrostatic potential $\delta v _{H} = v_H [\delta n]$, and the induced exchange-correlation potential 
\begin{equation}
\delta v_{\mathrm{xc}} = \int \mathbf{dr} \left.\frac{\delta v_{\mathrm{xc}}}{\delta n(\mathbf{r})}\right|_{n_0} n_{\mathrm{ind}} (\mathbf{r}).
\end{equation} 
The final result is
\begin{equation}
 \label{eq:TFexact}
 \int \mathbf{dr'} \left\lbrace  \left.\frac{ \delta ^2 T _s   [n]}{\delta n (\mathbf{r}) \delta n(\mathbf{r'})} \right| _{n_0}   n _{\mathrm{ind}}(\mathbf{r'}) \right\rbrace+ v_{\mathrm{scr}} (\mathbf{r}) = \delta \mu.
\end{equation}
Before showing how and under what conditions
Eq. (\ref{eq:TFcondition}) can be derived from Eq. (\ref{eq:TFexact}), it is worth mentioning that this derivation is equivalent to the standard linear response theory, in which the perturbation theory is applied to the self-consistent Kohn-Sham equations. \cite{zangwill80} 

Equation (\ref{eq:TFexact}) is an integral equation in which the exact form of the kernel is unknown.
The Thomas-Fermi approximation consists in approximating the kinetic functional to the free-electron gas case. Therefore, we have \cite{wang92}
\begin{eqnarray}
 \left.\frac{ \delta ^2 T _s [n]}{\delta n (\mathbf{r}) \delta n(\mathbf{r'})} \right| _{n_0}  &=&   \alpha \frac{10}{9} n_{0}(\mathbf{r})^{-1/3} \delta (\mathbf{r-r'}),
\end{eqnarray}
so that the integral condition Eq. (\ref{eq:TFexact}) reduces to the following algebraic expression:
\begin{equation}
\label{eq:TFcondition2} n_{\mathrm{ind}} (\mathbf{r}) = \frac{q (\mathbf{r})^2}{4\pi} \left[ \delta \mu - v_{\mathrm{scr}} (\mathbf{r})\right],
\end{equation}
where 
\begin{equation}
   q (\mathbf{r})= \left[\frac{9}{10 \alpha} n_0(\mathbf{r})^{1/3}\right]^{1/2}
\end{equation}
and $\alpha=3^{5/3}\pi^{4/3}/10$.
In the standard Thomas-Fermi model $n_0(\mathbf{r})$ is approximated as the constant spatial averaged valence electron density, which is the number of valence electrons in a unit cell divided by the unit cell volume. However, anticipating the results for Si and Ge to be discussed below, we have found that this approximation is not so important. What turns out to be crucial is the choice in Eq. (\ref{eq:TFexact}) for the second functional derivative of the kinetic energy. Indeed, writing it as a function of only the difference $\mathbf{r-r'}$ corresponds to neglecting the local field effects,
in that the response is independent of the coordinates in which an external point-charge impurity is located. This is equivalent, in a bulk system, to only considering the diagonal component of the reciprocal space dielectric matrix. This argument is about the kinetic functional, and does not regard the surface polarization effects, which are taken into account in the model through the use of suitable boundary conditions for the Poisson equation. We know that local field effects enter the theory through the kinetic functional derivative, as well as through the boundary conditions of the Poisson equation. The Thomas-Fermi theory neglects the first, but retains the second of such effects. In the bulk, only the first effect occurs, due to the polarization of bonds, and a link with the diagonal screening of the analysis of Ref. \onlinecite{fleszar87} can be traced.
The Thomas-Fermi theory gives results in fair agreement with the linear-response results, when the off-diagonal elements of the dielectric matrix are neglected.

In order to give a first indication on the validity of the Thomas-Fermi approximation in nanostructures, we compare the chemical potential $\delta \mu$ calculated from DFT with the Thomas-Fermi prediction. Within DFT, the chemical potential of a given structure is calculated as
\begin{equation}
 \mu = - \frac{(I + A )}{2},
 \label{eqmu}
\end{equation}
where $I=E^+_S - E_S$ and $A = E_S - E^-_S$  are, respectively, the structure ionization potential and the electron affinity ($E_S$ is the total energy).\cite{parr89} We have calculated the chemical potential variation $\delta \mu$ for Si$_{35}$H$_{36}$, for the isocoric P$^+$ impurity, obtaining $\delta \mu=-2.79$ eV from Eq. (\ref{eqmu}) and $\delta \mu=-2.62$ eV from the Thomas-Fermi Eq. (\ref{mudef}) (for the determination of the parameters, see below). Considering all the approximations involved, the agreement of less than $0.2$ eV can be considered very good.

\subsection{Model description and results} 
The Poisson equation, together with Eq. (\ref{eq:TFcondition}), gives the Thomas-Fermi equation for an external point-charge $Z$ located at the nanocrystal center,
\begin{equation}
 \label{eq:TFequation}
  \nabla ^2 v _{\mathrm{scr}} = -4\pi Z \delta(\mathbf{r})+q^2\left[ v_{\mathrm{scr}}(r) - \delta \mu \right].
\end{equation}
It can be written as 
\begin{equation}
 \label{eq:TFequation2}
  \nabla ^2 v_{\mathrm{scr}} = q^2\left[ v_{\mathrm{scr}}(r) - A \right]
\end{equation}
with the condition
\begin{equation}
 \lim _{r\to0} \left[ rv_{\mathrm{scr}}(r)\right] = -Z.
\end{equation}
The leading boundary condition is given by the continuity of the potential on the screening sphere,
\begin{equation}
   v_{\mathrm{scr}} (R_s) = v_c (R_s).
\end{equation}
Solving the Poisson equation, we find that the Thomas-Fermi expression for the effective spatial dielectric function, defined as the ratio between external and screened potential, is
\begin{widetext}
\begin{equation}
\tilde{\varepsilon} (r)=\left\{ 
\begin{array}{cl}
\left\lbrace \frac{1-Q/Z}{qR_s}\left\{\sinh\left[q(R_s-r)\right]+rq\right\} + \frac{r}{R} \frac{Q}{Z}\right\rbrace ^{-1}, & 0<r\leq R_s \\
\left\lbrace 1-\frac{Q}{Z} + \frac{r}{R}\frac{Q}{Z} \right\rbrace ^{-1}, & R_s \leq r \leq R. 
\end{array}
\right. \text{ \ }  
\label{eq:eps}
\end{equation}
\end{widetext}
Moreover, from the continuity of the electric field on the screening sphere, a relationship between the product $qR_s$ and $Q$ can be obtained,
\begin{equation}
\label{eq:resta}
   \frac{\sinh (qR_s)}{qR_s } = \frac{1}{1-Q/Z}.
\end{equation}
In the following, we shall use Eq. (\ref{eq:resta}) for determining the variable $q$ for each nanocrystal, once  the total density $Q$ and the screening radius $R_s$ are known.

The determination of the Thomas-Fermi parameters $Q$ and $R_s$ is not obvious. We propose to use the bulk $Q$, and calculate the screening radius from the \textit{ab initio} nanocrystal-induced density. As we shall see below, the $R_s$ dependence on the nanocrystal size is very weak, and, \textit{a posteriori}, we can say that the use of a constant value for the screening radius does not significantly change the results.
For determining $Q$ in the bulk, it is necessary to consider Eq. (\ref{eq:eps}) in the limit $R\to\infty$. It is easy to see that
\begin{equation}
\tilde{\varepsilon} _{bulk} (r)=\left\{
\begin{array}{cl}
\left\lbrace \frac{1-Q/Z}{qR_s}\left\{ \sinh\left[q(R_s-r)\right]+rq\right\} \right\rbrace ^{-1}, & 0<r\leq R_s \\
\left\lbrace 1-\frac{Q}{Z} \right\rbrace ^{-1}, & r \geq R_s.
\end{array}
\right. \text{ \ }
\label{eq:epsbulk}
\end{equation}
This is the equation first derived in Ref. \onlinecite{resta77}.
Since the material behaves like 
a dielectric medium beyond the screening radius, we get $ \varepsilon _s = 1/(1-Q/Z)$ from the above equation, in agreement with Eq. (\ref{eq:Qvseps0}). 

From this equation we have $Q=0.91$ and  $0.93$ for Si and Ge, respectively, corresponding to the bulk static dielectric constants $\varepsilon_s=11.4$ and $14.3$.
The choice for $Q$ in the case of Ge is motivated by the fact that it is the diagonal contribution to the integrated density, in the limit $r\to\infty$. \cite{fleszar87} Also, the corresponding $\varepsilon_s$ is very close to the Walter and Cohen pseudopotential calculations of dielectric function (they used $\varepsilon_s=14$ for bulk Ge). \cite{walter70} This allows us to show that the present model fits well to a standard RPA calculation. Finally, we want to point out here that the experimental value $\varepsilon_s =16$ is not far from the value we use. We have checked that small differences in the static dielectric constant do not significantly change our results.

It remains now to fix the screening radius $R_s$. This can be done starting from the DFT-induced charge illustrated in the previous section. One can define $R_s$ as the radius corresponding to an integrated induced charge equal to $Q$. Namely, the screening radius is given by the solution of the equation
\begin{equation}
   Q_{\mathrm{ind}} (R_s) = Q.
   \label{eq:rsdef}
\end{equation}
However, because of the wavy behavior shown by the induced charge, there can be several solutions to Eq. (\ref{eq:rsdef}).
Among the possible solutions, a choice can be done starting from physical motivations; in particular, we expect that the screening radius would take a value close to (actually slightly greater than) the impurity host bond length. Indeed, the whole displaced charge has to be included in the model, and it is likely that, after giving off the wavy contribution, the induced charge extends to the nearest-neighbor distance. \cite{fleszar87}

\begin{table}
\caption{Thomas-Fermi parameters calculated for several Si and Ge nanocrystals and for bulk Si. The nanocrystal radius $R$, the screening radius $R_s$, and the parameter $q$, as computed from Eq. (\ref{eq:resta}), are reported. All the quantities are in atomic units.}
\label{table1}
\begin{ruledtabular}
\begin{tabular}{cccc}
                       & $R$   & $R_s$ & $q$ \\
\hline
Si$_{35}$H$_{36}$        & 11.31 & 5.58 & 0.836   \\  
Si$_{87}$H$_{76}$        & 15.17 & 5.39 & 0.866  \\  
Si$_{147}$H$_{100}$      & 17.82 & 5.35 & 0.873  \\
Si$_{191}$H$_{148}$      & 19.58 & 5.36 & 0.871  \\
$Si_{bulk}$            & $\infty$ & 5.33  & 0.875\\
Ge$_{35}$H$_{36}$        & 11.92 & 5.90  & 0.84 \\
Ge$_{191}$H$_{148}$      & 20.62 & 5.61 & 0.882 
\end{tabular} 
\end{ruledtabular}
\end{table} 

In Table \ref{table1}, we report the calculated values for the Thomas-Fermi parameters, for several Si and Ge nanocrystals.
We have also included in this table the values for bulk Si that have been obtained with a calculation on a supercell comprising 512 atoms. The important result coming from Table \ref{table1} is that $R_s$ is quite independent of the nanocrystal size (apart from the case of Si$_{35}$H$_{36}$, very small variations are seen, being less than $0.1$ a.u. going from Si$_{87}$H$_{76}$ to the bulk Si). We can say that this independence of $R_s$ of the system dimension is a manifestation of the local nature of the response as originally suggested in Ref. \onlinecite{delerue03}.

It is important to point out that we use a different procedure from that in Ref. \onlinecite{resta77} in the determination of the parameters. In that case, $q$ was calculated starting from the material Fermi energy, while the screened radius was a derived variable, calculated from Eq. (\ref{eq:resta}). At variance with that procedure, we directly calculate the screening radius from first-principles results and then use Eq. (\ref{eq:resta}) to get $q$. The values of $R_s$ and $q$ obtained in Ref. \onlinecite{resta77} were $R_s=4.28$ and $q=1.10$ for silicon and $R_s=4.71$ and $q=1.08$ for germanium. Although the present estimations of $R_s$ ($q$) are bigger (smaller) than those of  Ref. \onlinecite{resta77}, we shall see below that the dielectric functions for both the bulk and nanocrystals are in excellent agreement with DFT and empirical pseudopotential results.

\begin{figure}
	\includegraphics[width=0.45\textwidth]{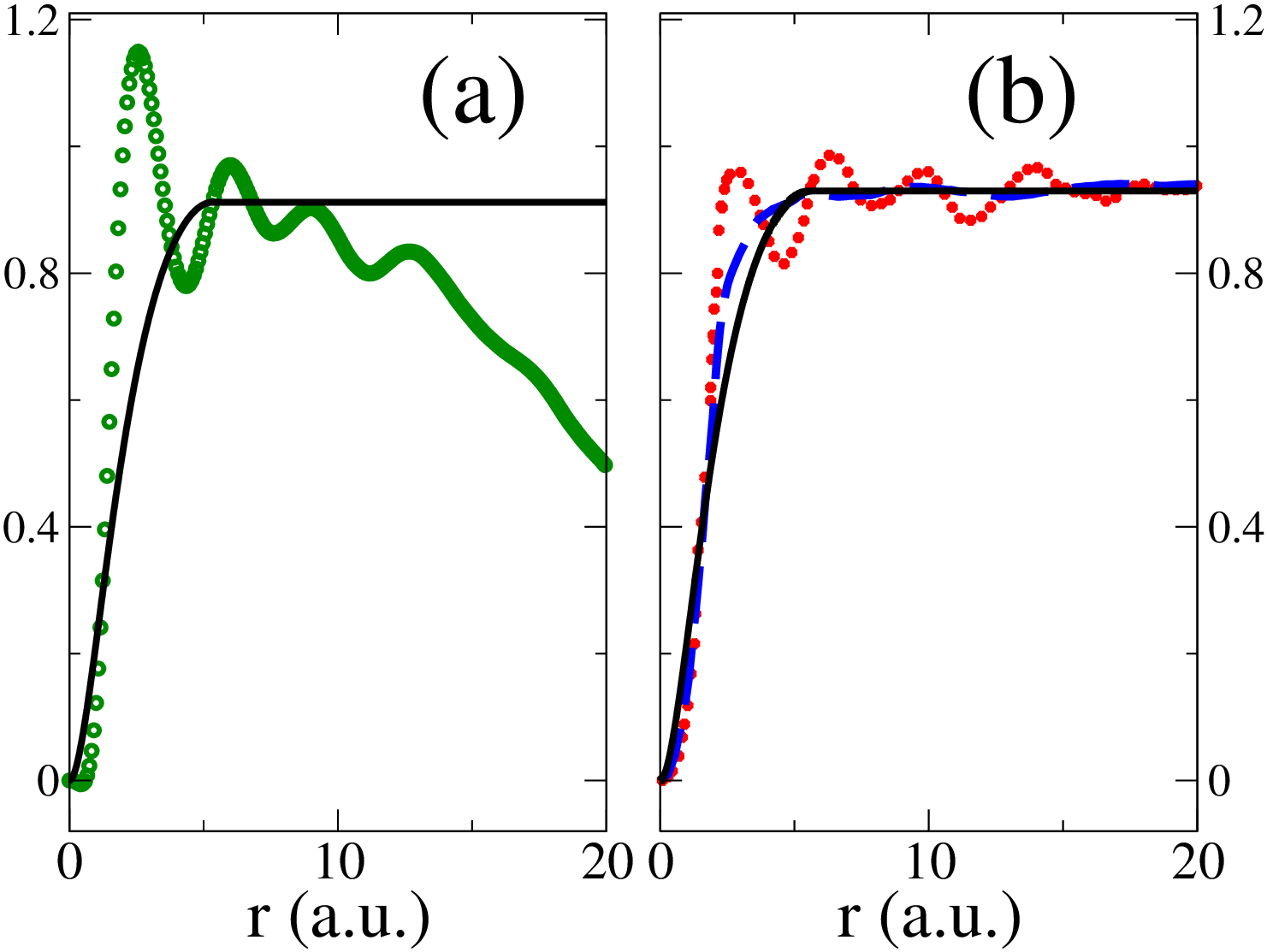}
	\caption{\label{fig:fig4}(Color online)	Integrated induced density calculated for bulk Si [panel (a)] and Ge [panel (b)] using both DFT and the Thomas-Fermi model. In panel (a), Thomas-Fermi (full, black line) and DFT results (circles, green) for Si$_{191}$H$_{148}$ nanocrystal.
	In panel (b), Thomas-Fermi (full, black line) and linear response results (stars, red) of Ref. \onlinecite{fleszar87}. The diagonal component of screening is also shown (dashed, blue line).}	
\end{figure}

Before presenting the results for the dielectric functions, let us look at the integrated densities in bulk silicon and germanium. The results are shown in Fig. \ref{fig:fig4}, where the solid line comes from the Thomas-Fermi model using, for silicon, the calculated bulk parameters and for germanium those corresponding to Ge$_{191}$H$_{148}$ (see Table \ref{table1}). In the case of Si [panel a], we compare the model with our nanocrystal DFT results. For Ge [panel b], a comparison is done with the linear-response calculation of Ref. \onlinecite{fleszar87}. We note from Fig. \ref{fig:fig4} that the characteristic wavy structure that is typical
of the full response (stars) is absent. More importantly, the agreement with our Thomas-Fermi calculation is very good. In other words, the model is perfectly capable of reproducing the response without local field effects. This is also true for silicon, as shown in Fig. \ref{fig:fig4}(a). Here it is clearly seen that the Thomas-Fermi curve lies in the middle of the DFT curve, as in the case of germanium. We can therefore reasonably conclude that the present model gives results consistent with an \textit{ab initio} calculation in which local field effects due to the bond polarization are either neglected or play a minor role.

\begin{figure}
	\includegraphics[width=0.5\textwidth]{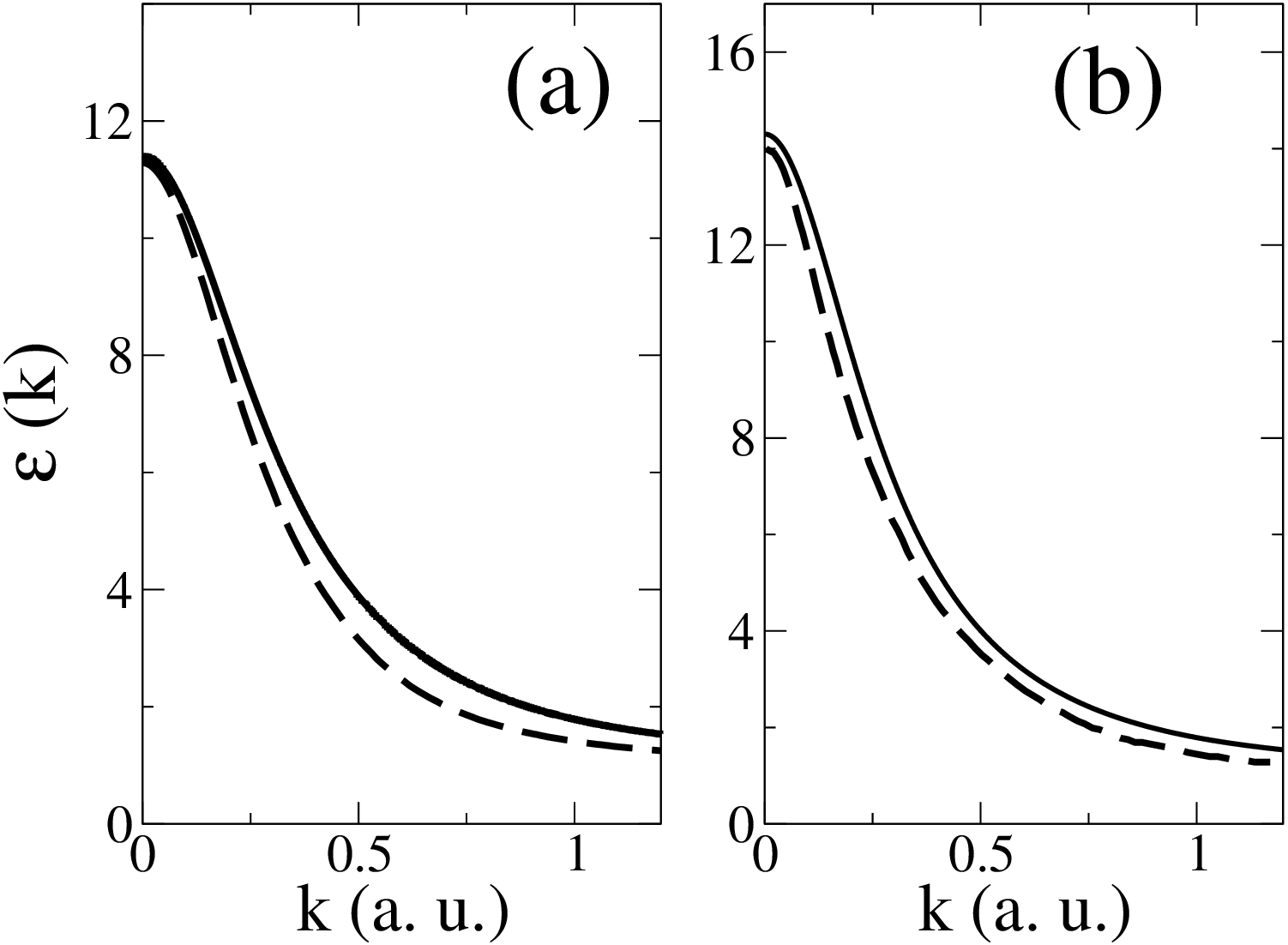}
	\caption{\label{fig:fig5}Wave-vector-dependent screening dielectric function for bulk Si [panel (a)] and bulk Ge [panel (b)]. The dashed lines are the RPA empirical pseudopotential calculation of Ref. \onlinecite{walter70}. The solid line are the Thomas-Fermi results from Eq. (\ref{epsk}).}	
\end{figure}

Within the present model, the wave-vector-dependent screening dielectric function of a bulk semiconductor takes the form
\begin{equation}
\varepsilon _s (k)=\frac{q^{2}+k^{2}}{q^{2}\left( 1-Q/Z\right) \sin
\left(kR_{s}\right) /kR_{s}+k^{2}}.
\label{epsk}
\end{equation}
In Fig. \ref{fig:fig5}, the results are shown for both bulk Si and Ge. A comparison with the empirical pseudopotential calculation of Walter and Cohen \cite{walter70} (dashed line) indicates that the model parameters listed in Table \ref{table1} give a very good agreement in the bulk limit.

\begin{figure}	
	\includegraphics[width=0.5\textwidth]{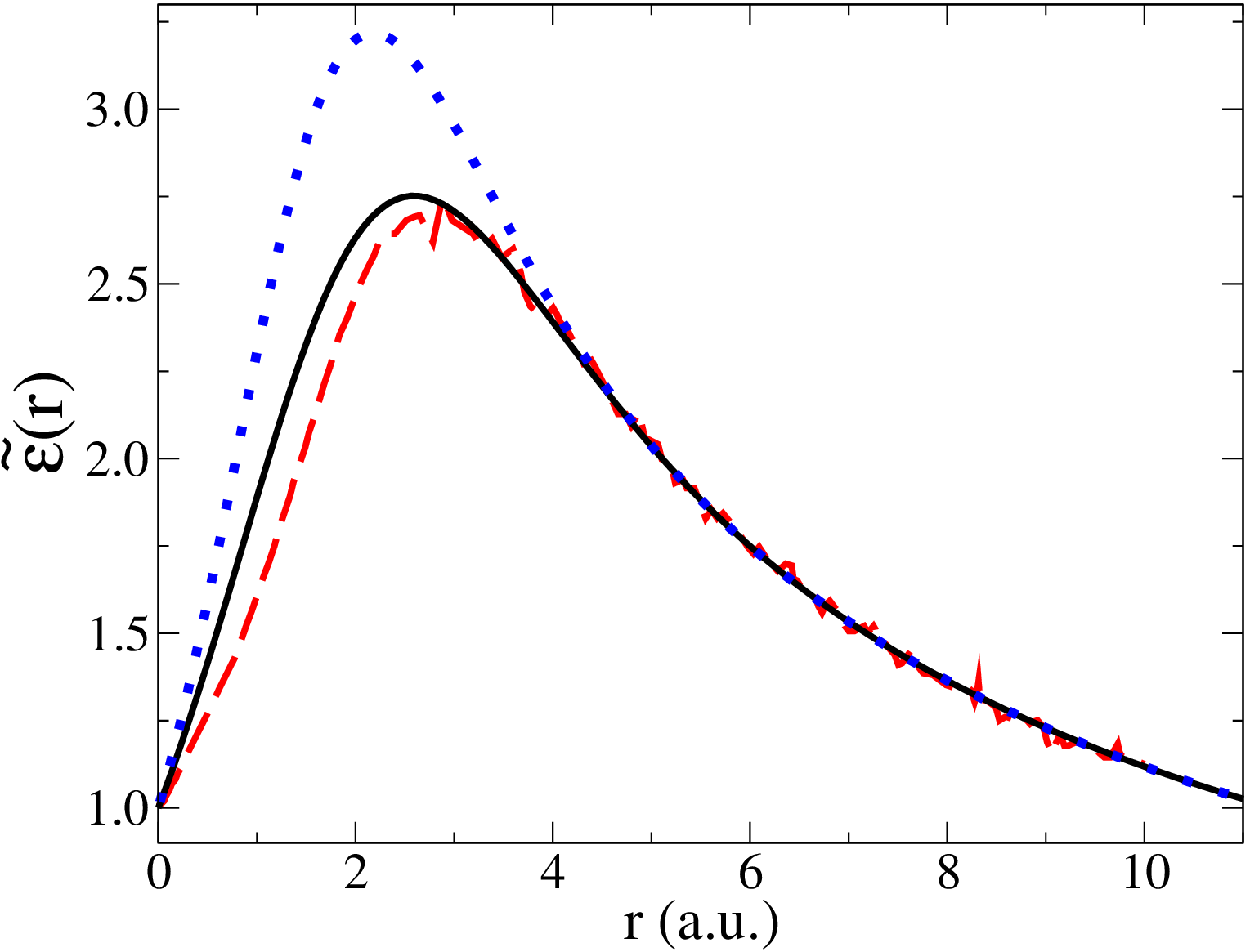}
        \caption{\label{fig:fig6}(Color online) Screening function $\tilde{\varepsilon} (r)$ for Si$_{35}$H$_{36}$. 
       Solid (black) line: Thomas-Fermi model with the parameters of Table \ref{table1};
       dotted (blue) line: the same but with the parameters of Ref. \onlinecite{resta77};
      dashed (red) line: real-space \textit{ab initio} calculation of Ref. \onlinecite{ogut03}.} 	
\end{figure}

In order to see how the present model performs in the case of nanocrystals, a comparison has been done with recent real-space \textit{ab initio} calculations. In Fig. \ref{fig:fig6}, the Si$_{35}$H$_{36}$ effective spatial screening function calculated with the Thomas-Fermi model and the \textit{ab initio} results taken from Ref. \onlinecite{ogut03} are shown.
For the Thomas-Fermi model, we have used both the derived values of $R_s$ and $q$ given in Table \ref{table1} (full line)
and those of Ref. \onlinecite{resta77} (dotted line). It is seen that in this last case the agreement with the \textit{ab initio}
results of Ref. \onlinecite{ogut03} is not perfect, particularly for the height and the position of the main peak. 
However, considering that there is no need of doing any calculation for getting the parameters of Ref. \onlinecite{resta77}, 
depending on the application at hand, this may surely be a first approximation. 
In any case, it is interesting to see that although the calculations of Ref. \onlinecite{ogut03} include the exchange and correlation terms, the agreement between our result (full line) and the \textit{ab initio} one (dashed line) is very impressive.
We want to remark once again that the most important part of the screening is due to electron charges accumulated both near the surface and close to the impurity.

We also performed a calculation of the donor impurity binding energy in order to compare the Thomas-Fermi model and DFT results.
Within a DFT framework, the binding energy can be calculated as the difference $E_{\mathrm{DFT}}=I_d-A_u$ between the doped nanocrystal ionization energy and the undoped nanocrystal electron affinity. \cite{melnikov04,cantele05} The binding energy can also be calculated
using the perturbation theory as 
\begin{equation}
\label{eq:Emdef}
E_m = -Z \left< \psi_c \left| \frac{1}{\tilde{\varepsilon}(r) r} \right|\psi_c \right>,
\end{equation}
where $\tilde{\varepsilon}(r)$ is the effective spatial dielectric function and $\psi_c$ is the first empty state of the undoped nanocrystal. Using the undoped nanocrystal DFT wave function $\psi_c$ and the effective dielectric function calculated with the model, we can compute the binding energy and compare it with the DFT results.
In Table \ref{table2}, the binding-energy results are shown. The interesting thing is that there is a systematic difference between the DFT results and the values estimated from the model. This size-independent contribution is quite small, about $0.2$ eV for Si. It can be due to both the bond polarization effects and the exchange and correlation contributions, which are not taken into account in the present model.

\begin{table}
\caption{Impurity activation energies calculated with both the \textit{ab initio} method ($E_{\mathrm{DFT}}$) and the Thomas-Fermi model ($E_{m}$) for several silicon nanocrystals. The DFT data are for the donor isocoric impurities P$^+$. All the values are in eV. }
\label{table2}
\begin{ruledtabular}
\begin{tabular}{cccccc}
                       &  $E_{\mathrm{DFT}}$ & $E_{m}$ \\
\hline
Si$_{35}$H$_{36}$        &  3.05 & 2.84 \\  
Si$_{87}$H$_{76}$        &  2.27 & 2.11 \\  
Si$_{147}$H$_{100}$      &  1.91 & 1.72 \\
Si$_{191}$H$_{148}$      &  1.79 & 1.60\\  
\end{tabular} 
\end{ruledtabular}
\end{table} 

\section{Conclusions}
Screening in covalent semiconductor nanocrystals has been studied using both advanced \textit{ab initio} methods and a Thomas-Fermi model. Our DFT calculations have shown that isocoric donor dopants essentially behave as a point-charge giving an induced charge that agrees well with the linear-response approximation. Comparing the induced integrated densities with those obtained in Ref. \onlinecite{fleszar87}, we have been able to conceptually isolate the features directly related to local field effects.
It is, at least in principle, not easy to distinguish the several contributions to the screening due to local fields. However, in the specific cases we have analyzed, there is an indication of the fact that surface polarization is the dominant local field effect in semiconductor nanocrystals. This was proved combining the Thomas-Fermi theory including electrostatics of surface polarization with the result that the screening function agrees very well with state-of-the-art \textit{ab initio} calculations.

It is worth mentioning that local field effects related to surface polarizations may have dramatic consequences also in the optical response, particularly for anisotropic structure such as wires and ellipsoidal dots. Indeed, it has been shown for both silicon \cite{bruneval05} and germanium \cite{bruno05} wires that the optical frequency-dependent absorption function is strongly suppressed for light polarized perpendicularly to the wire axis. This suppression does not come out when local fields are neglected. From a classical point of view, these local fields are again dominated by surface polarization, as originally recognized in interpreting the polarization anisotropy of porous silicon.\cite{kovalev} Although the present model cannot be
generalized to the case of the response to an external field, it is of interest to note that in both cases, that is, point-charge
screening and external field, surface polarization plays a fundamental role at the nanoscale.

The first-principles calculations presented in this work have shown that the concept of bulklike response to an external perturbation introduced in Ref. \onlinecite{delerue03} is valid also in a screening problem. Although the induced electron density contains some oscillations, we have shown that it is possible to define a screening radius that is practically independent of the nanocrystal dimension.
The progressive reduction of the screening action is due to the surface polarization. In this paper we have shown how this contribution may be described with simple electrostatics. The important point is that now we have an analytical expression for the effective screening function whose validity for both the nanoscale and bulk has been proved.

A final note should be devoted to a possible generalization of the model to the case of off-center impurities. This could be done, at least in principle, by considering the matching of the Thomas-Fermi solution to the full image potential of an off-center point-charge. However,  beyond the technical difficulties in doing that, one should consider that the method fails, particularly for very small nanocrystals, when the impurity
is at a distance from the nanocrystal surface that is less than or comparable to the screening radius.

\section*{ACKNOWLEDGMENTS}
We are grateful to Professor Serdar \"{O}\u{g}\"{u}t for a critical reading of the manuscript. Financial support by COFIN-PRIN 2005 is acknowledged. Calculations have been performed at CINECA-``Progetti Supercalcolo 2006'' and ``Campus Computational Grid''-Universit\`a di Napoli ``Federico II'' advanced computing facilities.


\end{document}